\documentclass[twocolumn,showpacs,preprintnumbers,amsmath,amssymb,aps]{revtex4}

\usepackage{graphicx}
\usepackage{amssymb}
\usepackage{bm}
\usepackage{epstopdf}
\begin{document}
    
\title{Measurement of the Neutron Lifetime Using a Proton Trap}
    
\author{M.~S.~Dewey, D.~M.~Gilliam, and J.~S.~Nico }
\affiliation{National Institute of Standards and 
Technology, Gaithersburg, MD 20899}

\author{F.~E.~Wietfeldt}
\affiliation{Tulane University, New Orleans, LA 70118}

\author{X.~Fei and W.~M.~Snow}
\affiliation{Indiana University, Bloomington, IN 47408}

\author{G.~L.~Greene} \affiliation{University of Tennessee/Oak Ridge
National Laboratory, Knoxville, TN 37996}

\author{J. Pauwels, R. Eykens, A. Lamberty, and J. Van Gestel}
\affiliation{European Commission, Joint Research Centre, Institute for Reference 
Materials and Measurements, 2440 Geel, Belgium}

\date{\today}

\begin{abstract}
We report a new measurement of the neutron decay lifetime by the
absolute counting of in-beam neutrons and their decay protons. Protons were
confined in a quasi-Penning trap and counted with a silicon detector.
The neutron beam fluence was measured by capture in
a thin $^{6}$LiF foil detector with known absolute efficiency. The combination of these
simultaneous measurements gives the neutron lifetime: 
$\tau_{n} = (886.8\pm1.2{\rm [stat]}\pm 3.2{\rm [sys]})$~s. The
systematic uncertainty is dominated by uncertainties in
the mass of the $^6$LiF deposit and the $^6$Li$(n,t)$ cross section. This is the most precise
measurement of the neutron lifetime to date using an in-beam method.
\end{abstract}

 
\maketitle

Precision measurements of neutron beta decay address basic questions in particle physics,
astrophysics, and cosmology. 
As the simplest semi-leptonic decay system, the free neutron plays a crucial role
in understanding the physics of the weak interaction and testing the validity of the Standard Model \cite{NDmot}.
The current experimental uncertainty in the neutron
lifetime dominates the uncertainty in calculating the primordial helium abundance, as a function of
baryon density, using Big-Bang nucleosynthesis \cite{BUR99}. Figure 1 shows a  summary of recent neutron lifetime 
measurements. The four most precise \cite{MAM89,NEZ92,MAM93,ARZ00} confined ultracold neutrons in a bottle; the decay lifetime
is determined by counting the neutrons that remain after some elapsed time,
with a correction for competing neutron loss mechanisms.  Two others
\cite{SPI88,BYR96} measured the absolute specific activity of a beam of cold neutrons by counting decay
protons. Given the very different systematic problems that the
two classes of experiments encounter, a more precise measurement of the
lifetime using the in-beam technique not only reduces the overall
uncertainty of $\tau_{n}$ but also provides a strong check on the
robustness of the central value. Note that the in-beam method measures only the neutron decay
branch that produces a proton in the final state, while the UCN bottle method measures all decay modes. With future
improvements in precision it may become possible to place important limits on exotic, nonbaryonic branches of neutron decay by comparing the lifetimes measured by the in-beam and UCN bottle methods. 

\begin{figure}
\includegraphics[width=8.0cm]{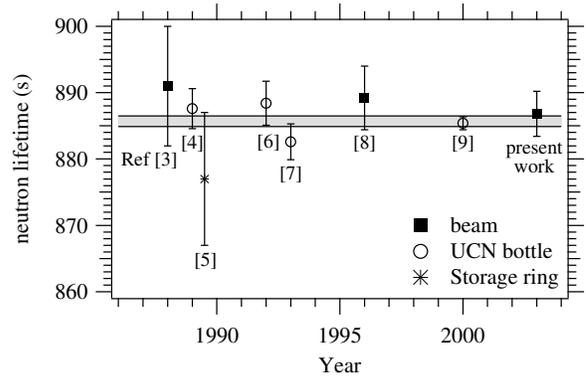}
\caption{Figure 1: A comparison of recent neutron lifetime measurements. The shaded band is $\pm 1$ standard deviation of the weighted average, including our result.}
\end{figure}

\par
Figure 2 shows a  sketch of the experimental method. It is based on the method first proposed by Byrne,  {\em et al.}, 
and is described in greater detail in previous publications \cite{PTmeth}.
A proton trap of length $L$ intercepts the entire width of the neutron beam.  Neutron
decay is observed by trapping and counting decay protons within the trap with an efficiency
$\epsilon_{p}$.  The neutron beam is characterized by a velocity
dependent fluence rate $I(v)$.  The mean number of neutrons inside the trap volume is
\begin{equation}
N_{n}= L \int_{A} da\, I(v) \frac{1}{v}
\label{eq:mean_n},
\end{equation}
where $A$ is the beam cross sectional area, 
and so the rate $\dot{N_{p}}$ at which decay protons are detected is
\begin{equation}
    \dot{N_{p}}=\frac{\epsilon_{p} L}{\tau_n}\int_{A}da\, I(v)\frac{1}{v}
\label{eq:mean_p}.
\end{equation}
After leaving the trap, the neutron beam passes through a
thin foil of $^6$LiF. The probability for absorbing a neutron in the foil
through the $^6{\rm Li}(n,t)^4$He reaction is inversely proportional to
the neutron velocity $v$. The reaction products, alphas or tritons, are counted by
a set of four silicon surface barrier detectors in a well-characterized geometry.
We define the efficiency for the neutron
detector, $\epsilon_{o}$, as the ratio of the reaction product rate to
the neutron rate incident on the deposit for neutrons with thermal velocity
$v_{o} = 2200$~m/s. The corresponding efficiency for neutrons of
other velocities is $\epsilon_{o}v_{o}/v$.  Therefore, the net reaction product
count rate $\dot{N_{\alpha}}$ is
\begin{equation}
    \dot{N_{\alpha}}=\epsilon_{o}v_{o}\int_{A}da\,I(v)\frac{1}{v}
\label{eq:mean_alpha}.
\end{equation}
The integrals in Eq.~(\ref{eq:mean_p}) and Eq.~(\ref{eq:mean_alpha})
are identical;  the velocity dependence of the neutron detector efficiency compensates for the fact that the
faster neutrons in the beam spend less time in the decay volume.  This cancellation is exact given two 
assumptions: (1) the neutron absorption efficiency in the $^6$LiF target is exactly proportional to $1/v$ and (2)
the neutron beam intensity and its velocity dependence do not change between the trap and the target. 
The deviation from the $1/v$ law in the $^6{\rm Li}(n,t)^4$He cross section at thermal and 
subthermal energies has been shown to be
less than 0.01\% \cite{Berg61}, and changes in the neutron beam due to decay in flight and residual gas interaction
are less than 0.001\%. Our only non-negligible correction to Eqs. (\ref{eq:mean_p}) and  (\ref{eq:mean_alpha}) 
is for the finite thickness of the $^6$LiF foil (+5.4 s). Thus we obtain the neutron lifetime $\tau_n$ 
from the experimental quantities $\dot{N_{\alpha}} / \dot{N_{p}}$, $\epsilon_{o}$, $\epsilon_{p}$, and $L$.

\begin{figure}
\vspace{-1.5cm}
\includegraphics[width=9.5cm]{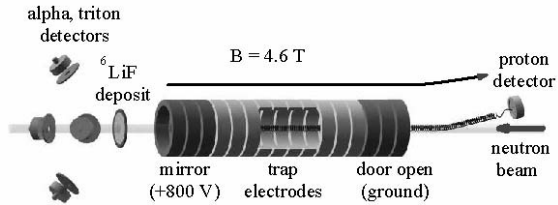}
\vspace{-2.0cm}
\caption{Figure 2: Schematic representation of the experimental method (not to scale).}
\end{figure}

\par
The experiment was performed at the cold neutron beamline NG6 at the National Institute of 
Standards and Technology (NIST) Center for Neutron Research. 
Cold neutrons were transported 68 m from the cold source to the experiment by a straight 
$^{58}$Ni-coated neutron guide. 
Immediately after exiting a local guide shutter,
the neutron beam passed through 20~cm to 25~cm of cooled single crystal bismuth to attenuate fast neutrons and gamma
rays that contribute to the background signal.  The beam was collimated by two
$^{6}$LiF apertures.  The diameter of the first aperture (C1)
was varied from 32~mm to 51~mm as a systematic check on the effect
the beam diameter on the measured lifetime.  The second (C2) had a
diameter of 8.4~mm and was not changed.  In between these two
beam-defining apertures were several $^{6}$LiF beam scrapers, which
removed scattered and very divergent neutrons. After C2, the
beam transited a 1~m section of pre-guide and then entered the vacuum
system through a 7.5~mm inner diameter quartz guide
tube.  After passing through the trap, it traveled 83~cm to the $^6$LiF deposit. The beam
exited the vacuum system
through a silicon window and stopped in a thick $^{6}$LiF beam dump. 
\par
The proton trap was a quasi-Penning trap, composed
of sixteen annular electrodes, each 18.6 mm long with an inner diameter of 26.0~mm, cut from fused quartz 
and coated with a thin layer of gold.
Adjacent segments were separated by 3 mm-thick insulating spacers of uncoated fused quartz.  
The dimensions of each electrode and spacer
were measured to a precision of $\pm5$~$\mu$m using a coordinate measuring machine at NIST.
Changes in the dimension due to thermal contraction are below the 10$^{-4}$ level for
fused quartz.
The trap resided in a 4.6 T magnetic field and the vacuum in the trap was maintained below $10^{-9}$~mbar.
\par
In trapping mode, the three upstream electrodes (the ``door'') were held
at +800 V, and a variable number of adjacent electrodes (the ``trap'') were held
at ground potential. The subsequent three adjacent electrodes (the ``mirror'')
were held at +800 V. We varied the trap length from 3 to 10 grounded electrodes. When a neutron
decayed inside the trap, the decay proton was trapped radially by the magnetic field 
and axially by the electrostatic potential in the
door and mirror. 
After some trapping period, typically 10 ms, the trapped protons were counted. 
The door was ``opened'', i.e. the door electrodes were lowered
to ground potential, and a small ramped potential was applied to the trap electrodes 
to assist the slower protons out the door. 
The protons were then guided by a 9.5$^{\circ}$ bend in the magnetic
field to the proton detector held at a high negative potential (-27.5~kV to -32.5~kV). 
After the door was open for 76 $\mu$s, a time
sufficient to allow all protons to exit the trap, the mirror was lowered to
ground potential. This prevented negatively charged particles, which may contribute to instability, from
accumulating in any portion of the trap. That state was maintained for 33 $\mu$s, after which the door and
mirror electrodes were raised again to +800 V and another trapping cycle began. 
Since the detector needed to be enabled only during extraction, the counting background was reduced by the
ratio of the trapping time to the extraction time (typically a factor
of 125). Figure 3 shows a plot of proton detection time for a typical run. 
\par
\begin{figure}
\includegraphics[width=8.0cm]{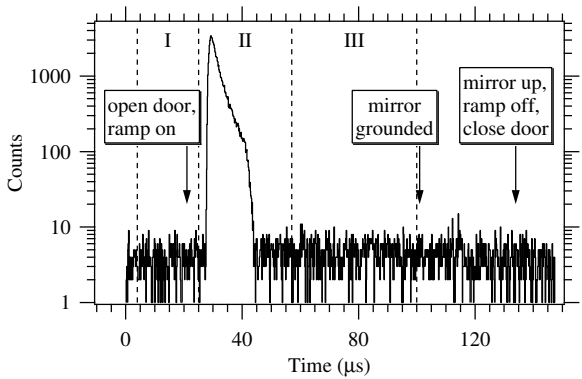}
\caption{Figure 3: A typical plot of proton detection time after gating on the detector ($t=0$). 
Regions I and III are used to subtract background from the peak region II. }
\end{figure}

Protons that were born in the trap (grounded electrode region) were trapped with 100\% efficiency. However protons
that were born near the door and mirror (the ``end regions''), where the electrostatic potential is elevated, 
were not all trapped. A proton born in the end region was trapped if its initial (at birth) sum of 
electrostatic potential energy and axial kinetic energy was less than the maximum end potential. 
This complication caused the effective length $L$ of the trap to be 
difficult to determine precisely. It is for this reason that we varied the trap length. 
The shape of the electrostatic potential near the door and mirror 
was the same for all traps with 3--10 grounded electrodes, so 
the effective length of the end regions, while unknown, was in principle constant. 
The length of the trap can then be written $L = nl + L_{\rm end}$
where $n$ is the number of grounded electrodes and $l$ is the physical length of one electrode plus
an adjacent spacer. $L_{\rm end}$ is an {\em effective} length of the 
two end regions; it is proportional to the physical length of the end regions {\em and} the 
probability that protons born there will be trapped. From Eq. (\ref{eq:mean_p}) and (\ref{eq:mean_alpha})
we see that the ratio of proton counting rate to alpha counting rate is then
\begin{equation}
\frac{\dot{N_p}}{\dot{N}_{\alpha}} = \tau_n^{-1}\left( \frac{\epsilon_p}{\epsilon_0 v_o}\right)
(nl + L_{\rm end}).
\end{equation}
We fit $\dot{N_p} / \dot{N_{\alpha}}$ as a function of $n$ to a straight line and determine $\tau_n$ from
the slope, so there is no need to know the value of $L_{\rm end}$, provided that
it was the same for all trap lengths. 
\par
Because of the symmetry in the Penning trap's design, $L_{\rm end}$ was approximately
equal for all trap lengths. There were two trap-length-dependent effects that broke the
symmetry: the gradient in the axial magnetic field, and the divergence of the neutron beam. Each of these effects caused $L_{\rm end}$
to vary slightly with trap length. 
A detailed Monte Carlo simulation of the experiment, based on the measured and calculated magnetic and electric field inside the trap, was developed in order to correct for these trap nonlinearities. 
It gave a trap-length dependent correction that lowered the lifetime by 5.3 s.
\par
Surface barrier (SB) and passivated
ion-implanted planar silicon (PIPS) detectors were used at various times for counting the protons.  
The proton detectors were large enough so that all protons produced by neutron decay in the collimated beam, 
defined by C1 and C2, would strike the 19.7-mm diameter active region after the trap was opened. 
The detector was optically aligned to the magnetic field axis,
and the alignment was verified by scanning with a low energy electron source at the trap's center and with actual neutron decay protons.
When a proton hit the active region of the detector, the efficiency for proton detection was less than unity because:  
1) a proton can lose so much energy in the inactive surface dead layer that it deposits insufficient 
energy in the active region to be detected;  2) a proton can be stopped within
the dead layer and never reach the active region; 3) a proton can backscatter from the dead layer or the active region and deposit
insufficient energy. In the first two cases, the proton will definitely not be detected. We denote the fraction of protons lost
in this manner by $f_{\rm Lost}$.  In the last case, there is some probability that the backscattered proton will be reflected back 
to the detector by the electric field and subsequently be detected. 
We denote the fraction of protons that backscatter but still have some chance for detection by $f_{\rm Bsc}$.
\par
To determine the proton detection efficiency, we ran
the experiment with a variety of detectors with different dead layer thicknesses and different acceleration potentials.
The fraction of protons lost $f_{\rm Lost}$ and the fraction that backscatter $f_{\rm Bsc}$ were calculated using the SRIM 2000 Monte
Carlo program \cite{ZIE00}. We found that
$f_{\rm Lost}$ varied from $4.0(3)\times 10^{-5}$ to $8.0(6)\times 10^{-3}$, and $f_{\rm Bsc}$ from $1.83(13)\times 10^{-3}$ to $2.37(17)\times 10^{-2}$, depending on the detector and acceleration potential.
The proton counting rate $\dot{N_p}$ for each run was multiplied by $1 + f_{\rm Lost}$ to correct for lost protons. The correction for backscattered protons was not as simple because of the unknown probability for a backscattered proton to return and be detected, so we made an extrapolation of the measured neutron lifetime to zero backscatter fraction (see Figure 4).
\par
The neutron detector target was a thin (0.34~mm), 50-mm-diameter single crystal
wafer of silicon coated with a 38~mm diameter deposit of $^6$LiF, fabricated at the Institute
for Reference Materials and Measurements (IRMM) in Geel, Belgium. 
The manufacture of deposits and characterization of the $^{6}$LiF areal density were exhaustively detailed in
measurements performed over several years~\cite{TAG91}.  The average areal density was $\rho = (39.30 \pm
0.10)~\mu\rm{g/cm^{2}}$. The $\alpha$ particles and tritons produced by the
the $^6{\rm Li}(n,t)^4$He reaction were detected
by four surface barrier detectors, each with a well-defined and carefully measured
solid angle. The geometry was chosen to make the solid angle subtended by the detectors insensitive to first order in the source position.
The parameter $\epsilon_{0}$ gives the ratio of detected alphas/tritons to incident thermal neutrons. 
It was calculated using
\begin{equation}
\epsilon_{0} =\frac{\sigma_{0}}{4\pi}
\int\int\Omega(x,y)\rho(x,y)\theta(x,y) dx dy \label{eq:neutroneff},
\end{equation}
\noindent where $\sigma_{0}$ is the cross section at thermal ($v_{0} =
2200$~m/s) velocity, $\Omega(x,y)$ is the detector solid angle,
$\rho(x,y)$ is the areal mass density of the deposit, and
$\theta(x,y)$ is the areal distribution of the neutron intensity on
the target. The $^{6}$Li thermal cross section is
($941.0 \pm 1.3$)~b ~\cite{CAR93}. It is important to note that we take the Evaluated Nuclear Data Files
(ENDF/B-6) $1\sigma$ uncertainty from the evaluation, {\em not} the expanded
uncertainty, to be the most appropriate for use with this precision experiment. 
The neutron detector solid angle was measured in two independent ways: mechanical contact
metrology and calibration with $^{239}$Pu alpha source of known
absolute activity. These two methods agreed to within 0.1~\%.
\par
The neutron beam intensity distribution was measured at both ends of the trap, and at the neutron monitor, using the
Dy foil imaging technique \cite{CHO00}. This method provides almost 5 decades of dynamic range, making it ideal for sensitive neutron measurements. These quantitative images were used to determine the beam distribution function $\theta(x,y)$ in Eq. \ref{eq:neutroneff}. They were also used to measure the beam ``halo'', the scattered neutrons that lie outside the collimated neutron beam. Trapped decay protons that originate from the halo may have trajectories that lie outside the active radius of the proton detector so they are not counted, while these halo neutrons are counted by the neutron counter. We found that 0.1~\% of the neutron intensity is in the halo outside the effective radius of the proton detector.
\par

\begin{figure}
\includegraphics[width=8.0cm]{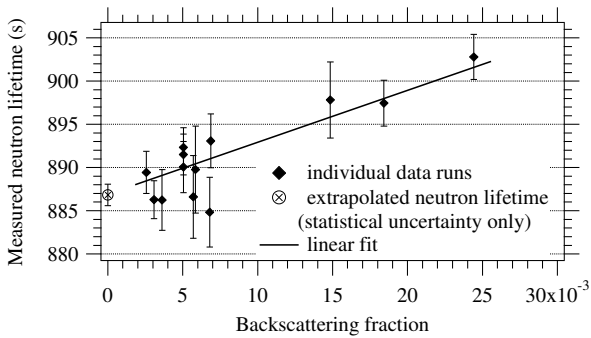}
\caption{Figure 4: A linear fit of the measured neutron lifetime at varying values of the detector backscattering fraction.  The extrapolation to zero backscattering gives the free neutron lifetime.}
\end{figure}

Proton and neutron counting data were collected for 13 run series, each with a different proton detector and acceleration potential. The corrected value of the neutron lifetime for each series was calculated and plotted vs. backscattering fraction, as shown in Figure 4. A linear
extrapolation to zero backscattering gave a result of $\tau_{n} = (886.8\pm1.2{\rm [stat]}\pm 3.2{\rm [sys]})$~s. A summary of all corrections and uncertainties is given in Table \ref{tab:Errors}.  This result would be improved by an independent absolute calibration of the neutron counter, which would significantly reduce the two largest systematic uncertainties, in the $^{6}$LiF foil density and $^{6}$Li cross section.  A cryogenic neutron radiometer that promises to be capable of such a calibration at the 0.1\% level has recently been demonstrated \cite{Zema}, and we are pursuing this method further. With this calibration, we expect that our experiment will ultimately achieve an uncertainty of less than 2 seconds in the neutron lifetime.
\par
We thank  R.~D.~Scott for assistance in the
characterization of the $^{6}$LiF and $^{10}$B targets. We also thank J.~Adams, J.~Byrne, A.~Carlson, Z.~Chowdhuri, P.~Dawber, C.~Evans, G.~Lamaze, and P.~Robouch
for their very helpful contributions, discussions, and
continued interest in this experiment. We gratefully acknowledge the support of NIST
(U.S. Department of Commerce), the U.S. Department of Energy (interagency agreement No.\ DE-AI02-93ER40784) and the National Science Foundation (PHY-9602872).

\begin{table}
\caption{\label{tab:Errors} Summary of systematic corrections and 
uncertainties (in seconds) for the measured neutron lifetime.}
\begin{ruledtabular}
\begin{tabular}{lcc}
Source 										&Correction &Uncertainty\\
\hline
$^{6}$LiF foil areal density	 					&   	      &2.2\\
$^{6}$Li cross section							&   	      &1.2\\
Neutron detector solid angle 						&+1.5      &1.0\\
Neutron beam halo								&-1.0      &1.0\\
LiF  target thickness								&+5.4      &0.8\\
Trap nonlinearity 								&-5.3     &0.8\\
Neutron losses in Si wafer						&+1.3     &0.5\\
$^{6}$LiF distribution in deposit 				         & -1.7      &0.1\\
Proton backscatter calc.				  			&             &0.4 \\
\hline
Proton counting statistics							&    	     &1.2\\
Neutron counting statistics						&   	     &0.1\\
\hline
Total											& +0.2      &3.4\\
\end{tabular}
\end{ruledtabular}
\end{table}

\end{document}